\documentclass{PoS}
\usepackage{amsmath,amssymb}
\title{Zero modes of Overlap fermions, instantons, and monopoles}

\ShortTitle{Zero modes of Overlap fermions, Instantons, and Monopoles}

\author{Adriano Di Giacomo\\
        University of Pisa, Department of Physics, Largo B. Pontecorvo, 3, Pisa, 56127, Italy\\
        E-mail: \email{digiaco@df.unipi.it}}

\author{\speaker{Masayasu Hasegawa}\\
        Joint Institute for Nuclear Research, Bogoliubov Laboratory of Theoretical Physics, Dubna, Moscow, 141980, Russia\\  
        (Previous affiliation: University of Parma, Department of Physics and INFN, Via G. P. Usberti 7/A, Parma, 43124, Italy)\\
        E-mail: \email{hasegawa@theor.jinr.ru}\\}

\abstract{The purpose of this study is to investigate the relations between instantons, monopoles, and Chiral symmetry breaking. The monopoles are important topological configurations existing in QCD which are believed to produce colour confinement. The groups of University of Kanazawa and Pisa have produced by Lattice simulations many results supporting the idea that QCD vacuum is a dual superconductor. Instantons are related to Chiral symmetry breaking, as explained e.g. in the instanton liquid model of E. V. SHURYAK. Clarifying quantitatively the relation between monopoles and instantons is not easy, also because monopoles are three dimensional objects, while instantons are four dimensional. We generate configurations, adding monopole-antimonopole pairs of opposite charges by a monopole creation operator. We observe that the monopole creation operator only adds long monopole loops in the configurations. Then, we count the number of fermion zero modes in the configurations using Overlap fermions as a tool. Finally, we find that one monopole-antimonopole pair makes one zero mode of plus or minus chirality, that is to say, one instanton of plus or minus charge.}

\FullConference{The 32nd International Symposium on Lattice Field Theory,\\
		23-28 June, 2014\\
		Columbia University New York, NY}

\begin{document}

\begin{normalsize}
\section{Introduction}

We carry out simulations to show the relations between zero modes of Overlap fermions, instantons, and monopoles as follows:\\
\noindent \textbf{(1) Overlap fermions}\\
First, we generate quenched configurations of the Wilson gauge action, construct Overlap operator from gauge links, solve the eigenvalue problem, and finally, count the number of fermion zero modes in the configuration.\\
\noindent  \textbf{(2) The number of Instantons}\\
We count the number of instantons from the number of zero modes. However, we never observe coexistence of zero modes of opposite chirality in the same configuration. We always observe zero modes of only chirality plus \textbf{or} only chirality minus in all configurations. Nevertheless, we assume that this is probably due to a kind of instability of pairs of opposite zero modes at small volumes in their detection, so that the topological charge is correct. Indeed the instanton density calculated from the average square of the topological charges is consistent with the result of the instanton liquid model~\cite{Shuryak1}.\\
\noindent \textbf{(3) Additional monopoles}\\
In order to clarify quantitatively the relation between the monopoles and instantons, one monopole-antimonopole pair with opposite charges is directly added in the configurations by a monopole creation operator. The monopole creation operator is defined in~\cite{DiGiacomo1, DiGiacomo2, DiGiacomo3}.\\
\noindent \textbf{(4) Measuring the additional monopoles}\\
We check whether the pair of monopoles is successfully added in the configurations or not, by use of the method developed in~\cite{Kanazawa2}. We find that adding one monopole-antimonopole pair makes one long monopole loop in configurations.\\
\noindent \textbf{(5) The relations between Zero modes, instantons and monopoles}\\
Lastly, we generate configurations adding one pair of one monopole and one anti-monopole with charges ranging from zero to four. We count the number of zero modes for each value of the monopole charge by the Overlap Dirac operator, and calculate the average square of topological charge. Finally, we compare analytic predictions based on our assumption on absence of opposite sign zero modes with the simulation results. 
\section{Overlap fermions}

In the numerical computations~\cite{Galletly1}, the massless Overlap Dirac operator $D(\rho)$ is defined as follows:
\begin{equation}
D(\rho) = \frac{\rho}{a} \left[ 1 + \frac{D_{W}(\rho)}{\sqrt{D_{W}(\rho)^{\dagger}D_{W}(\rho)}}\right], \ \ D_{W}(\rho) = D_{W} - \frac{\rho}{a}, \ (\rho = 1.4). 
\end{equation}
$\rho$ is a (negative) mass parameter $0 < \rho < 2$. $D_{W}$ is the massless Wilson Dirac operator defined as follows:  
\begin{align}
&D_{W} = \frac{1}{2}[\gamma_{\mu}(\nabla_{\mu}^{*} + \nabla_{\mu}) - a\nabla_{\mu}^{*}\nabla_{\mu}]\\
& [\nabla_{\mu}\psi](n) = \frac{1}{a}[U_{n,\mu}\psi(n+\hat{\mu}) - \psi(n)], \ [\nabla_{\mu}^{*}\psi](n) = \frac{1}{a}[\psi(n) - U_{n-\hat{\mu}, \mu}^{\dagger}\psi(n-\hat{\mu})]
\end{align}
There are exact zero modes of plus chirality $n_{+}$ and minus chirality $n_{-}$ in eigenvalues of this massless Overlap Dirac operator. The topological charge is defined as $Q = n_{+} - n_{-}$, the susceptibility $\langle Q^{2} \rangle$/V is computed from the topological charges.\\ 
The massless Overlap Dirac operator is calculated by the sign function using the Chebyshev polynomial approximation as follows:
\begin{equation}
\frac{D_{W}(\rho)}{\sqrt{D_{W}(\rho)^{\dagger}D_{W}(\rho)}} = sgn(D_{W}(\rho)) \equiv \gamma_{5}sgn(H_{W}(\rho)), \ H_{W}(\rho) = \gamma_{5}D_{W}(\rho).
\end{equation}
$H_{W}(\rho)$ is Hermitian Wilson Dirac operator of $D_{W}(\rho)$. We use this $H_{W}(\rho)$ operator for computations of a minmax polynomial approximation~\cite{Giusti1}.

\subsection{Simulation details}

We generate configurations using the Wilson gauge action. The numbers of configurations we use in simulations are $\mathcal{O}(200) \sim \mathcal{O}(800)$ for each value of the parameters $\beta$ and Volume, a total of 17 choices for the parameters. The Overlap Dirac operator is constructed from gauge links of the configurations. The eigenvalue problems $D(\rho)|\psi_{i}\rangle = \lambda_{i}|\psi_{i}\rangle$ are solved by the subroutines of ARPACK, and $\mathcal{O}$(80) pairs of the low-lying eigenvalues $\lambda_{i}$ and eigenvectors $|\psi_{i}\rangle$ are saved. The index $i$ is the pair number ($1 \leqq i \leqq \mathcal{O}(80)$). In our simulations, the lattice spacing is calculated following~\cite{Necco1}, and the Sommer scale $r_{0} = 0.5 \ [\mbox{fm}]$ is used.

\subsection{The number of Zero modes, the topological charges, and the topological susceptibility}

\begin{figure}    
\begin{minipage}[t]{0.45\textwidth}
\includegraphics[width=75mm]{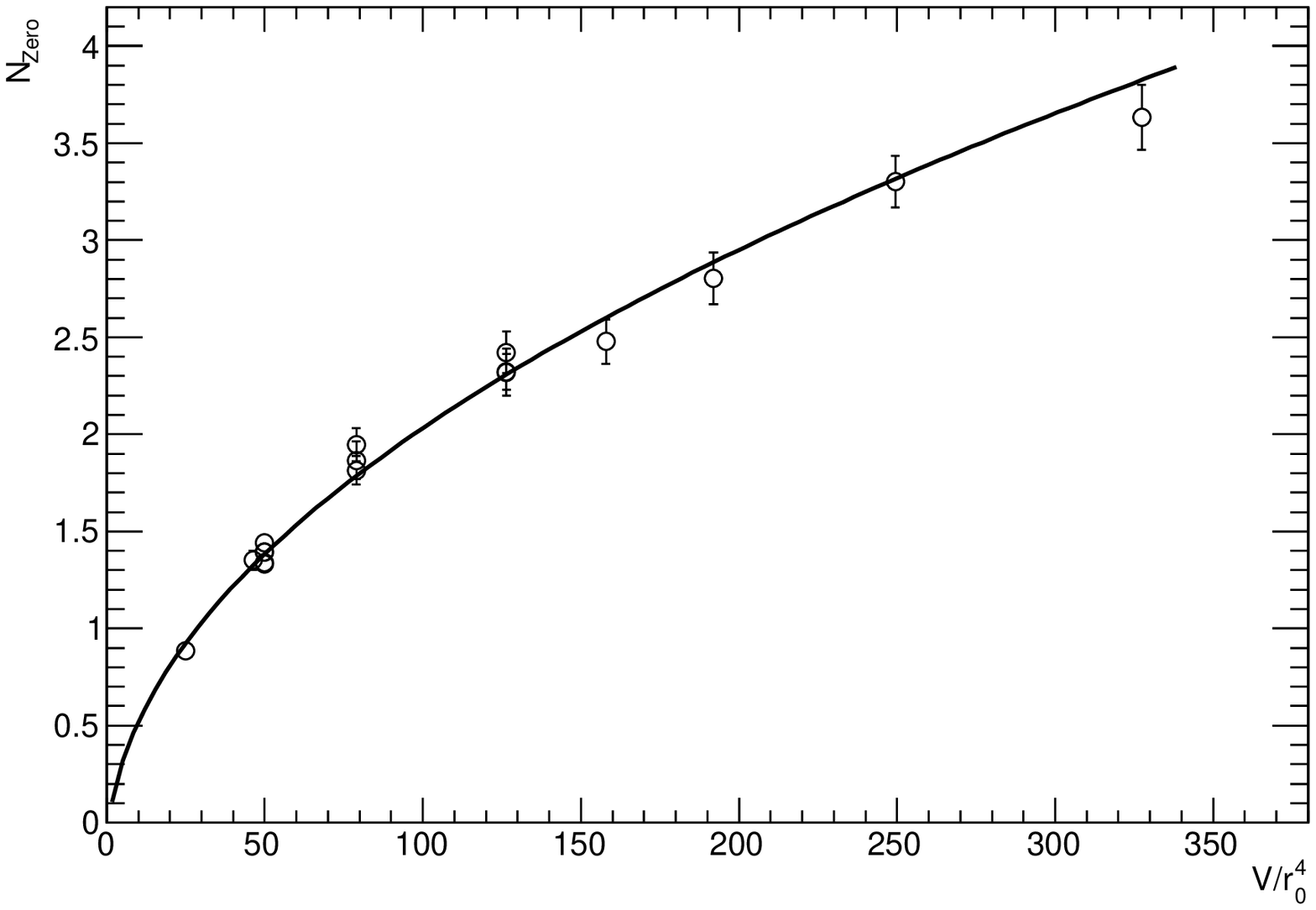}
\caption{The observed number of zero modes $N_{Zero}$ vs the physical volume $V/r_{0}^{4}$.}
\label{fig:Zero_allV1}
\end{minipage}
\hspace{\fill}
\begin{minipage}[t]{0.45\textwidth}
\includegraphics[width=75mm]{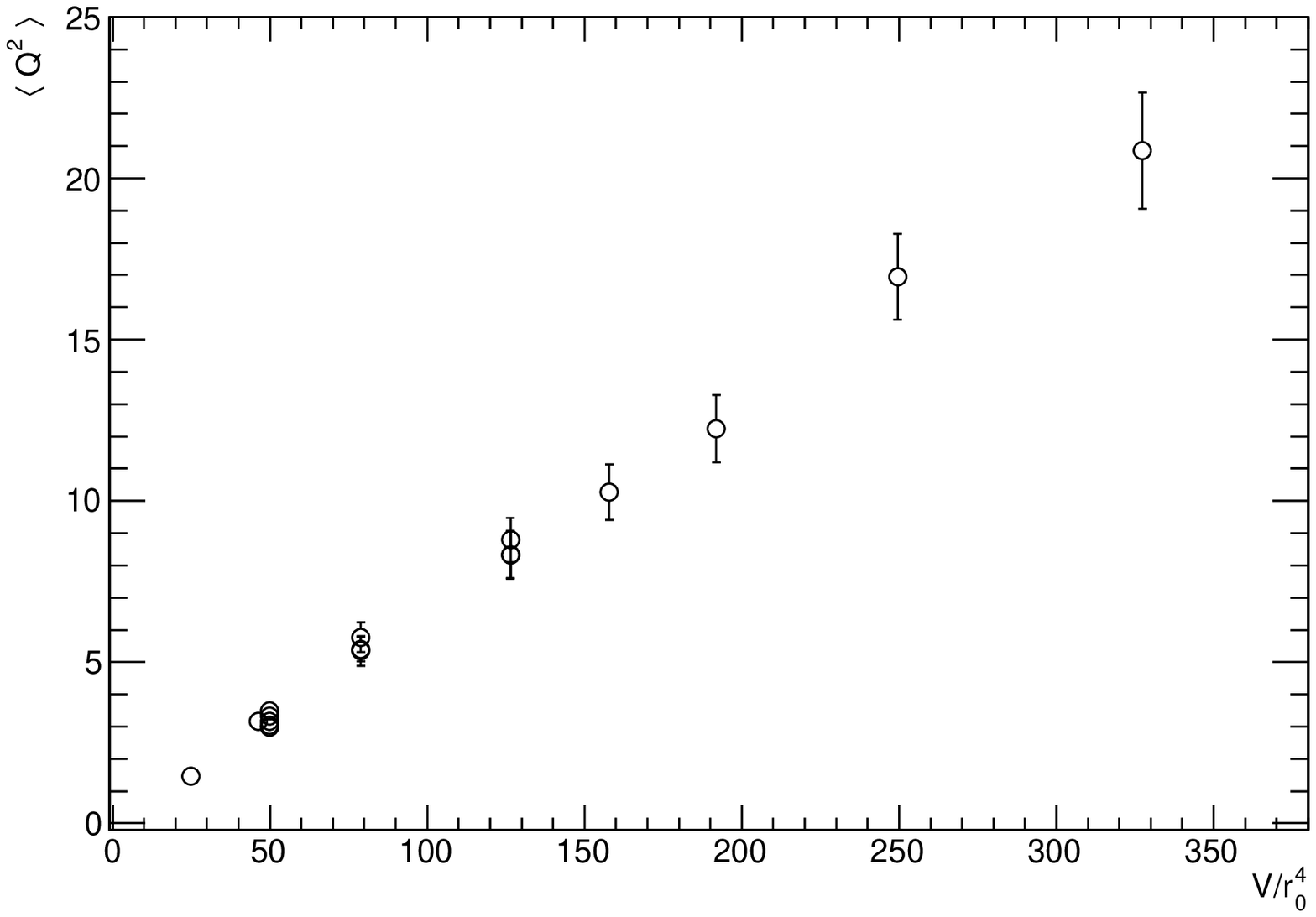}
\caption{The average square of topological charges $\langle Q^{2} \rangle$ vs the physical volume $V/r_{0}^{4}$.}
\label{fig:Q2_allv_2}
\end{minipage}
\end{figure}
\begin{figure}    
\begin{minipage}[t]{0.45\textwidth}
\includegraphics[width=75mm]{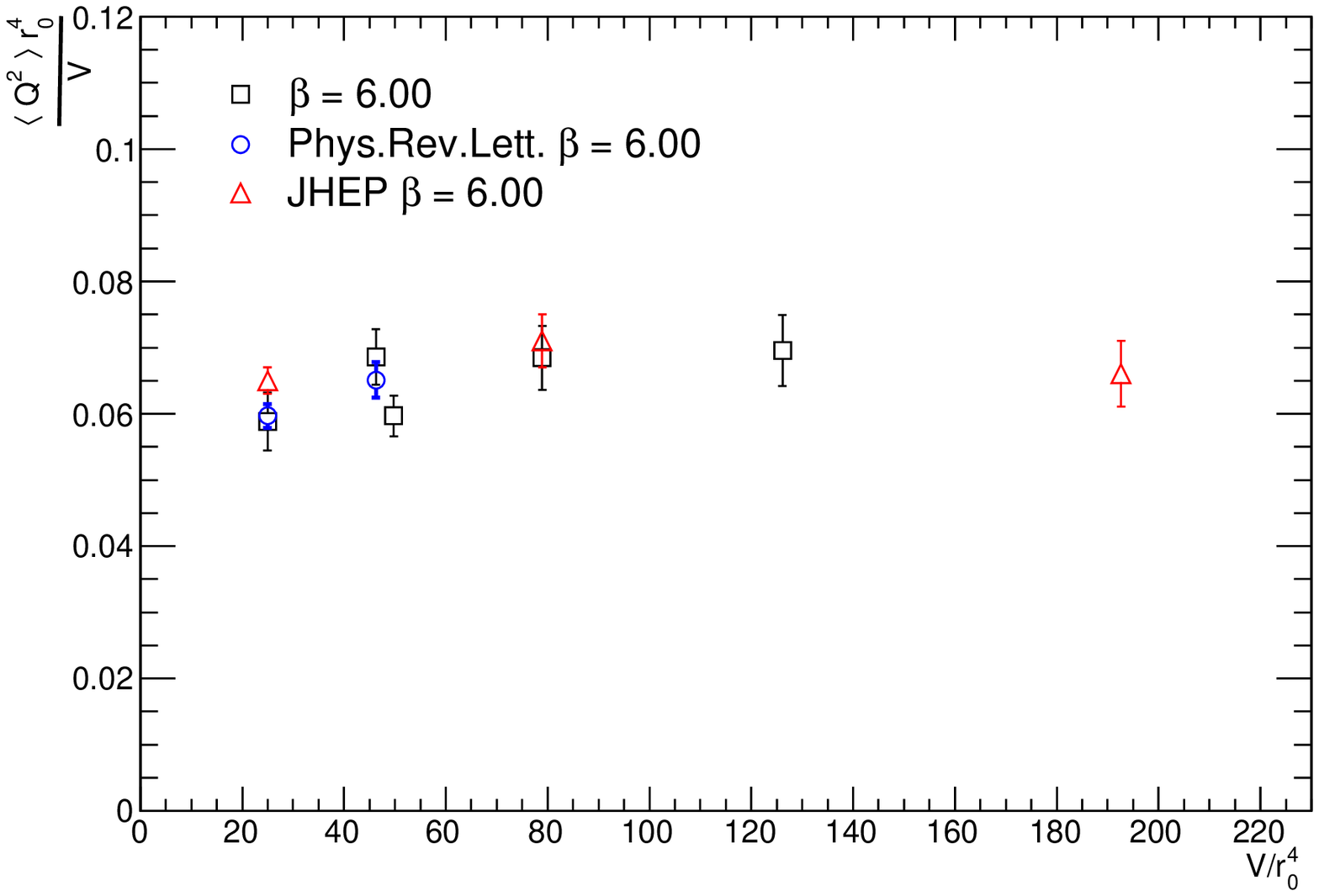}
\caption{The topological susceptibility of $\beta = 6.00$ together with other group results~\cite{Giusti2, Giusti3}.}
\label{fig:Tops_sus_b6p00}
\end{minipage}
\hspace{\fill}
\begin{minipage}[t]{0.45\textwidth}
\includegraphics[width=75mm]{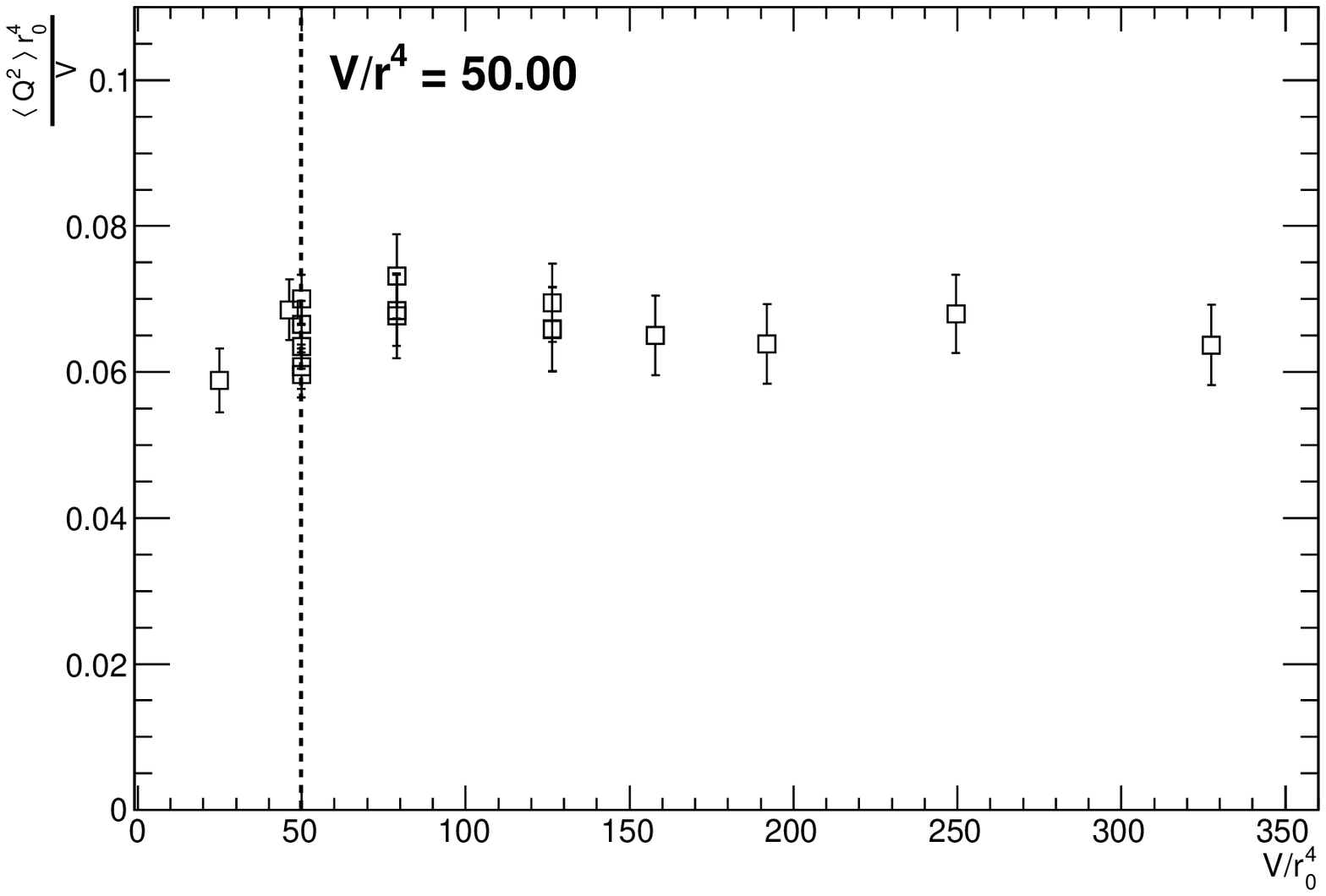}
\caption{The all results of the topological susceptibilities in our simulations vs the physical volume. A dotted line indicates the physical volume $V/r_{0}^{4} = 50.00$.}
\label{fig:Top_sus_all1}
\end{minipage}
\end{figure}

In our simulations, we never observed zero modes of + chirality and zero modes of - chirality in the same configuration. The zero modes in our simulation have only + chirality or only - chirality in each configuration. The observed number of zero modes $N_{Zero}$ increases with the physical volume $V/r_{0}^{4}$ as shown in Figure \ref{fig:Zero_allV1}. We suppose that the number of zero modes we observe is the \textbf{net} number of zero modes ($n_{+} - n_{-}$), that is to say the topological charges Q. The average square of topological charges $\langle Q^{2}\rangle$ is proportional to the physical volume $V/r_{0}^{4}$ as shown in Figure \ref{fig:Q2_allv_2}. We compare the topological susceptibility with results of other groups in Figure \ref{fig:Tops_sus_b6p00}. The figure shows that our results are consistent with them. Moreover, we check that the finite lattice volume does not affect the topological susceptibility up to $V/r_{0}^{4} = 327.4$ ($L = 2.1$ [fm]), Figure \ref{fig:Top_sus_all1}. 

\subsection{Topological susceptibility in the continuum limit}\label{sec:T_sus_con}

Last, we fix the physical volume at $V/r_{0}^{4} = 50.00$ indicated in Figure \ref{fig:Top_sus_all1}, and extrapolate the five data points of the topological susceptibility to the continuum limit using a linear expression $\langle Q^{2}\rangle r_{0}^{4}/V = c_{0} + c_{1} a^{2}$. Here, we compare with other groups. The results are consistent. Therefore, we confirm that eigenvalues and eigenvectors of overlap fermions in our simulations are properly computed.
\begin{align}
& \mbox{\textbf{Our result:}} \ \chi  = (1.86 (6) \times 10^{2} \ \ [\mbox{MeV}])^{4} \\
& \mbox{Ref~\cite{DelDebbio1}:} \ \chi = (1.88 \pm 12 \pm 5\times 10^{2} \ \ [\mbox{MeV}])^{4}\\
& \mbox{Ref~\cite{Giusti3}:} \ \chi  = (1.91 (5) \times 10^{2} \ \ [\mbox{MeV}])^{4}\\
& \mbox {The theoretical expectation ~\cite{Veneziano1, Witten1}:} \ \chi =\frac{\mbox {F}_{\pi}}{6} (m^{2}_{\eta} + m^{2}_{\eta'} - 2m_{\mbox{K}}^{2})|_{exp} \simeq (1.80\times 10^{2} \ \ [\mbox{MeV}])^{4}
\end{align}

\section{Instantons}\label{sec:inst}

 We assume that we observe the \textbf{net} number of zero modes, because physical lattice volumes are too small to distinguish $n_{+}$ and $n_{-}$ in one configuration. Then, the zero modes in our simulations are defined as $\pm N_{\pm} \equiv n_{+} - n_{-} = \pm Q$, and also the number of zero modes in this study is defined as $N_{Zero} = |Q|$ = $N_{+} : (n_{+} - n_{-} > 0)$, $0  : (n_{+} - n_{-} = 0)$, or $N_{-} : (n_{+} - n_{-} < 0)$. The number of instantons is determined as $N = \langle Q^{2} \rangle = \langle N_{Zero}^{2}\rangle$. The result for the instanton density is 
\begin{equation}
\rho_{i} = 8.3 (3) \times 10^{-4} \ [\mbox{GeV}^{4}].
\end{equation}

\section{The monopole creation operator}\label{sec:mon_cre1}

The monopole creation operator is defined in Ref.~\cite{DiGiacomo1, DiGiacomo2, DiGiacomo3}. Specifically, in this study, we use the monopole creation operator defined in~\cite{DiGiacomo3}. 
The monopole creation operator is defined as
\begin{equation}
\mu = \exp(- \beta \Delta S)
\end{equation}
$\Delta S$ is defined by modifying the normal plaquette action $S$  as follows:
\begin{equation} 
S + \Delta S = \sum_{n, \mu < \nu} \mbox{Re} (1 - \bar{\Pi}_{\mu\nu}(n))
\end{equation}
$\bar{\Pi}_{i0}$ is as a modified plaquette which is inserted matrices $M_{i}(\vec{n})$ and $M_{i}(\vec{n})^{\dagger}$ below, 
\begin{equation} 
\bar{\Pi}_{i0}(t, \vec{n}) = \frac{1}{\mbox{Tr}\mbox[I]}\mbox{Tr}[U_ {i}(t, \vec{n})M_{i}^{\dagger}(\vec{n} + \hat{i})U_{0}(t, \vec{n} + \hat{i})M_ {i}(\vec{n} + \hat{i})U_{i}^{\dagger}(t + 1, \vec{n})U_{0}^{\dagger}(t, \vec{n})].
\end{equation}
The matrix $M_{i}(\vec{n})$ 
\begin{equation} 
M_{i}(\vec{n}) = \exp(i m_{c} A_{i}^{0}(\vec{n} - \vec{x})), \ ( i = x, \ y, \ z ).
\end{equation}
is the discretised form of the classical field configuration $A_{i}^{0}(\vec{n} - \vec{x})$ produced by the monopoles. We take a monopole-antimonopole pair of charges $\pm m_c$, with
\begin{equation}
m_{c} = 0, 1, 2, 3, 4.\label{eq:def_mcharge1}
\end{equation}
The monopole has charge $+m_{c}$ and the anti-monopole charge $-m_{c}$, and the total monopole charges is zero. The monopole $+m_{c}$ sits in (x, y, z) the anti-monopole $-m_{c}$ in (x', y', z') at the certain time slice T, and at a given distance in the lattice. In Monte Carlo simulations pairs of monopoles make long monopole loops in the configurations. 

\section{Measuring the additional monopoles}\label{sec:ma_mono1}

 \begin{figure}[htbp]
  \begin{center}
 \includegraphics[width=75mm]{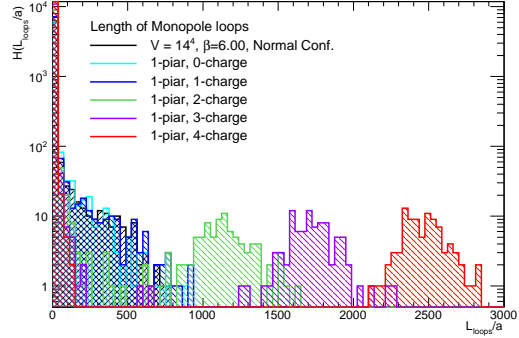}
 \caption{A histogram of the length of the monopole loops. The monopoles with charges are added in the configurations. The monopole charges are increased from 0 to 4.}
 \label{fig:Mon_loops_hist1}
  \end{center}
 \end{figure}

To check that the monopoles are successfully added, we detect the monopoles in the configurations. It has been found that the monopoles are divided in two clusters~\cite{Kanazawa3, DeGrand1, Kanazawa2, Hart1} in MA gauge. The small (ultraviolet) clusters are composed of the short monopole loops. The large (infrared) clusters which percolate through the lattice and wrap around the boundaries of lattice are formed by the longest monopole loop $L_{loops}$. The method of numerical computations of the monopole world line in four dimension is explained in~\cite{Bode1}. If the physical lattice volume is large enough, the small clusters and the large clusters are separated. We measure the length of monopole loops, and make a histogram as in Figure \ref{fig:Mon_loops_hist1}. We find that the monopole creation operator makes only long monopole loops in vacuum, and that the length of monopole loops increases with the number of monopole charges.

\section{The relations between Zero modes, instantons and monopoles}

\begin{figure}    
\begin{minipage}[hb]{0.45\textwidth}
\includegraphics[width=71.0mm]{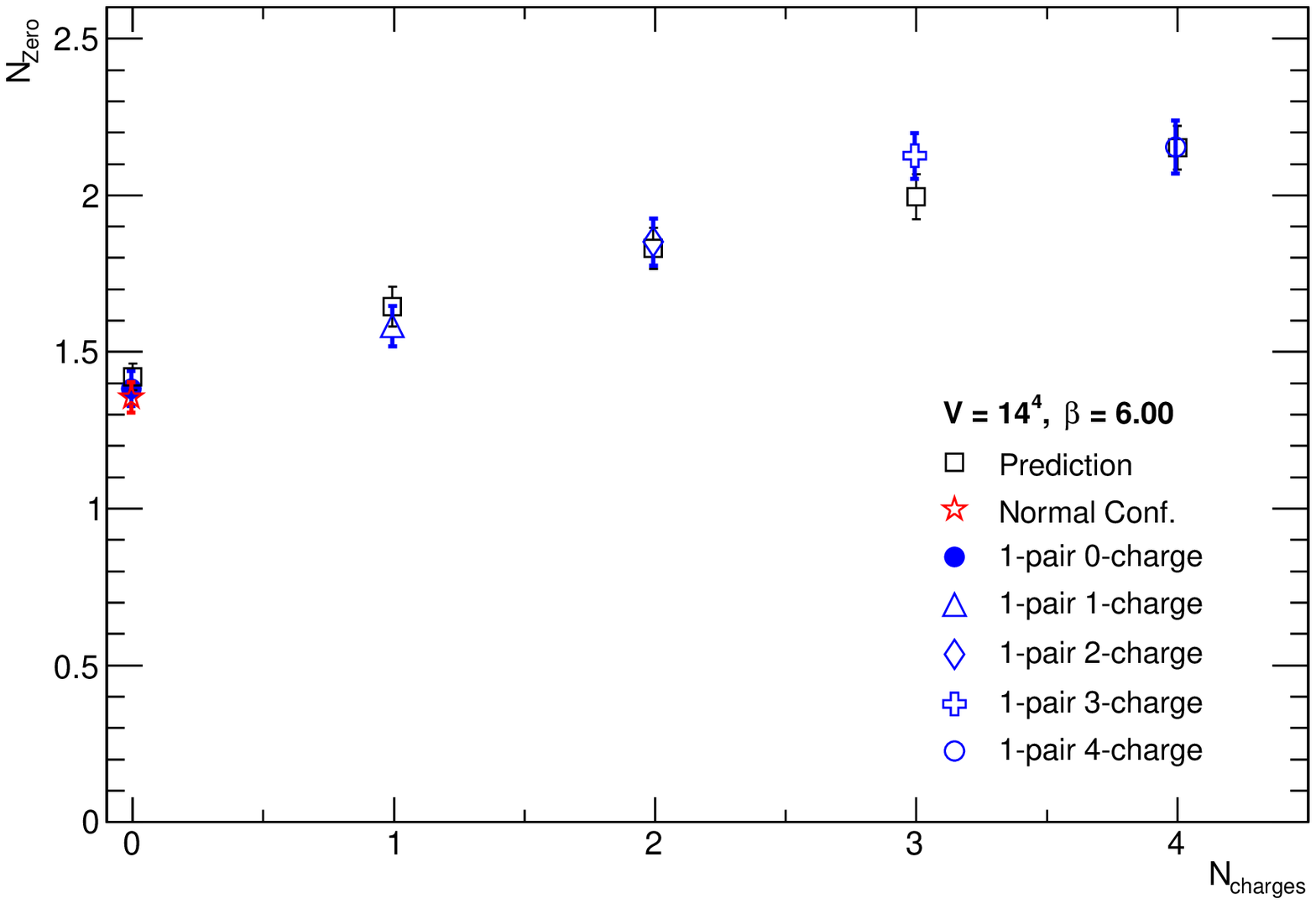}
\caption{The number of zero modes vs the number of monopole charges.}
\label{fig:Add_zero1}
\end{minipage}
\hspace{\fill}
\begin{minipage}[hb]{0.45\textwidth}
\includegraphics[width=71.0mm]{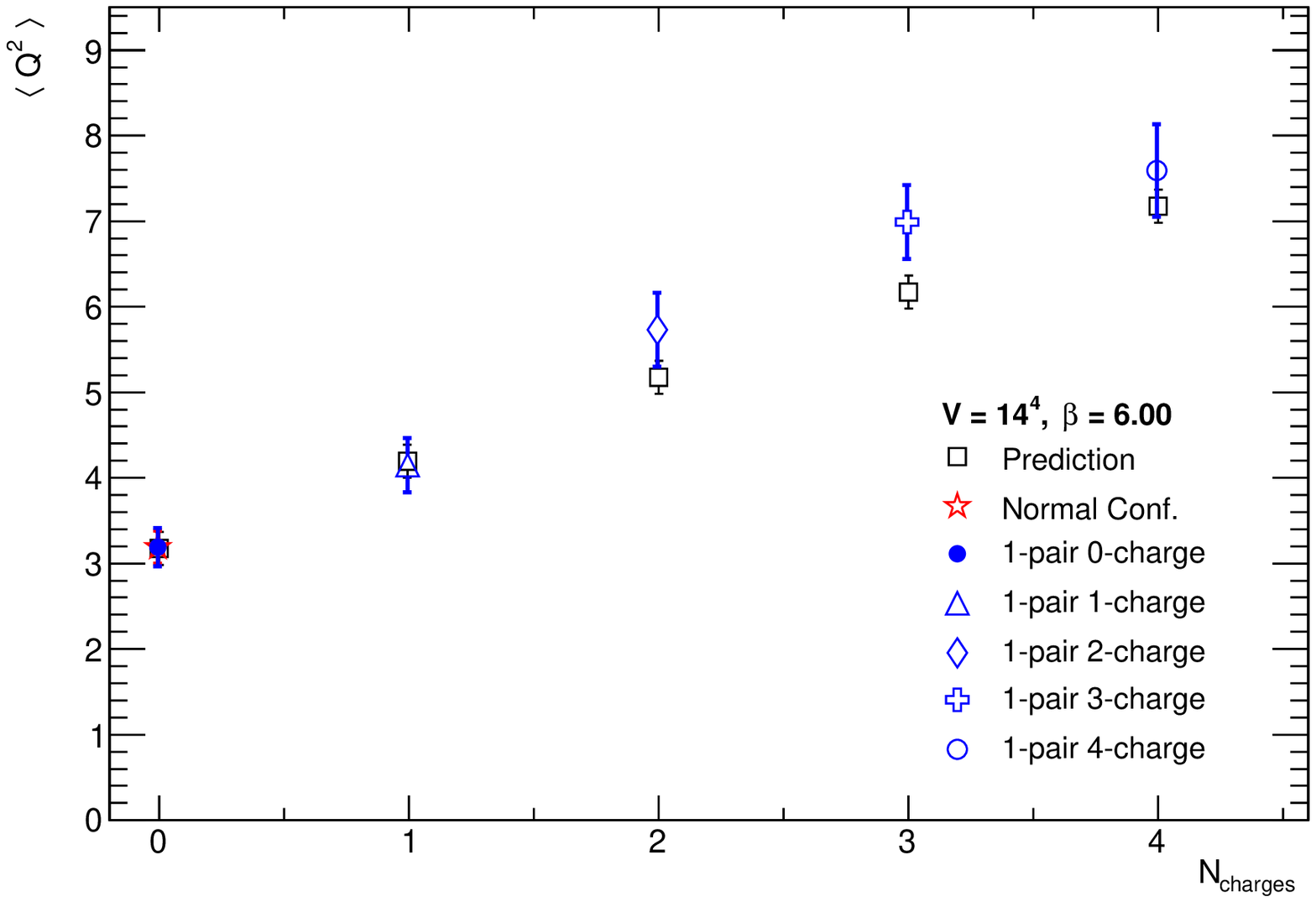}
\caption{The average square of topological charges vs the number of monopole charges.}
\label{fig:Add_q21}
\end{minipage}
\end{figure}
 We generate configurations with monopole-antimonopole pair varying the magnetic charges of the monopoles $m_{c}$ ($N_{charges}$) from zero to four. The distances between the monopole and anti-monopole are fixed at 6, and 8. The numbers of configurations are used as follows: $\mathcal{O}(200) \sim \mathcal{O}(300)$, for the distance 6; $\mathcal{O}(400)$, for the distance 8. We count the numbers of zero modes $N_{Zero}$ in the spectrum of eigenvalues, and calculate the average square of topological charges $\langle Q^{2} \rangle$. We compare the simulation results with an analytic prediction as indicated in Figure \ref{fig:Add_zero1} and Figure \ref{fig:Add_q21}. The distance is 8. We do not do smearing, cooling, or MA gauge fixing in the simulations.
\begin{table}[htb]
\begin{center}
\begin{small}
\begin{tabular}{|c|c|c|c|c|} \hline
      & A & B & Fit Range ($N_{charges}$)      &$\chi^{2}/d.o.f.$  \\ \hline 
  Prediction   &  1.00  &  3.17 (19)   & - & - \\ 
  Distance 6     &  1.02 (13) &  2.90 (19) & 0 - 4 & 7.9/3.0 \\
  Distance 8     &  1.19 (11) &  3.1 (2)   & 0 - 4 & 1.4/3.0 \\ \hline
\end{tabular}
\caption{The final results. The slope A that is computed by the analytical prediction is exactly 1. The intercept B by the analytical prediction is $\langle Q^{2} \rangle = 3.17(19)$. This value is computed from normal configurations.}\label{tb:f_res1}
\end{small}
\end{center}
\end{table}
 Concerning the analytic prediction in the figures, first, we suppose that one pair of one monopole with plus one charge and one anti-monopole with minus one charge makes one zero mode of plus chirality or minus chirality in one configuration. However, we can not observe the zero mode by our simulations. Instead of the zero mode, we observe the topological charge $Q = n_{+} - n_{-}$ as discussed in the Section \ref{sec:inst}. Then we analytically calculate the average square of topological charges, when one pair of monopoles with 0, 1, 2, 3, and 4 charges are added in configurations. We fit a linear function y = Ax + B to the simulation results to check the consistency. The final results are listed in Table \ref{tb:f_res1}. Those results are consistent. That is to say that the one monopole with plus one charge and one anti-monopole with minus one charge make one instanton. 

\section{Summary and Conclusion}

We discussed the number of zero modes, topological susceptibility in the continuum limit, the number of instantons, adding monopoles into configurations, measuring the monopole loops in the configurations, and the relation between the number of zero modes and the charges of the monopoles. The number of instantons is directly proportional to the physical volume. The instanton density is consistent with the instanton liquid model by E. V. SHURYAK. We confirm that one pair of one monopole and one anti-monopole is successfully added in configurations by the monopole creation operator, by measuring the length of the long monopole loops. We find that one monopole with plus one charge and one anti-monopole with minus one charge make one instanton. 

\section{Acknowledgments}
M. H. really appreciates help by E.-M. Ilgenfritz, Y. Nakamura, F. Pucci, T. Sekido, and V. Weinberg. M. H. thanks the University of Bielefeld, University of Kanazawa, University of Parma, and University of Pisa for hospitality. This research is supported by the Research Executive Agency (REA) of the European Union under Grant Agreement No. PITN-GA-2009-238353 (ITN STRONGnet). M. H. receives partial supports from the Istituto Nazionale di Fisica Nucleare at the University of Parma, and University of Pisa. We thank to the Research Center for Nuclear Physics and to the Cybermedia Center at the University of Osaka, and the Yukawa Institute at the University of Kyoto for the computer time and the technical support.

\end{normalsize}
\end{document}